\documentclass[aps,pra,showpacs,floatfix,superscriptaddress,twocolumn]{revtex4}
\usepackage[dvips]{graphicx}

\usepackage{longtable}
\usepackage{dcolumn}
\usepackage[dvips]{graphicx}
\usepackage{bm}
\usepackage{bbm}

\usepackage{times}
\usepackage{nicefrac}
\usepackage{amsmath}
\usepackage{amsfonts}
\usepackage{amssymb}
\usepackage{amsthm}

\newcolumntype{.}{D{x}{}{-1}}

\newcommand{\vare}{\varepsilon}

\newcommand{\lbr}{\langle}
\newcommand{\rbr}{\rangle}

\newcommand{\SixJ}[6]{
        \left\{
        \begin{array}{ccc}
        #1  & #2  & #3 \\
        #4  & #5  & #6 \\
        \end{array}
        \right\}
        }

\begin{document}

\title{Relativistic
configuration-interaction calculation of $\bm{K\alpha}$ transition energies in beryllium-like iron}

\author{V. A. Yerokhin}

\affiliation{Helmholtz-Institut Jena, D-07743 Jena, Germany}
%\affiliation{Friedrich-Schiller University Jena, D-07743 Jena, Germany}

\affiliation{Center for Advanced Studies, St.~Petersburg State Polytechnical University,
 195251 St.~Petersburg, Russia}

\author{A. Surzhykov}

\affiliation{Helmholtz-Institut Jena, D-07743 Jena, Germany}

\author{S. Fritzsche}

\affiliation{Helmholtz-Institut Jena, D-07743 Jena, Germany}
\affiliation{{Theoretisch-Physikalisches Institut, Friedrich-Schiller-Universit\"at Jena, D-07743
Jena, Germany}}

\begin{abstract}

We perform relativistic configuration-interaction calculations of the energy levels of the
low-lying and core-excited states of beryllium-like iron, Fe$^{22+}$. The results include the QED
contributions calculated by two different methods, the model QED operator approach and the
screening-potential approach. The uncertainties of theoretical energies are estimated
systematically. The predicted wavelengths of the $K\alpha$ transitions in beryllium-like iron
improve previous theoretical results and compare favourably with the experimental data.

\end{abstract}


\maketitle

\section{Introduction}

Highly charged iron produces some of the brightest X-ray emission lines from hot astrophysical
objects, such as compact X-ray binaries, galaxy clusters and stellar coronae. The $K\alpha$
spectral features of iron have been detected in the spectra of nearly all classes of cosmic X-ray
sources, because of their high transition rate, low intergalactic absorption as well as the high
relative abundance of iron in the Universe. Moreover, iron has been found an important element for
the diagnostics of hot laboratory plasma, notably in magnetic nuclear fusion and tokamaks. Since
the Fe $K\alpha$ line typically contains the contributions from different charge states, its
analysis provides useful information about the equilibrium and non-equilibrium charge-state
distributions of ions as well as about the electron and ion temperatures in the plasma.

In view of this importance of the Fe $K\alpha$ line for astrophysics and laboratory diagnostics,
accurate theoretical predictions are needed for the reliable identification and interpretation of
experimental spectral data. The simplest ion contributing to the $K\alpha$ line is the helium-like
ion. {\em Ab initio} QED calculations are available for these ions \cite{artemyev:05:pra}, whose
accuracy is significantly higher than the present experimental precision. For more complicated
ions, however, no full-scale QED calculations of  $K\alpha$ transitions have yet been performed,
and one has to rely on some kind of relativistic calculations complemented by an approximate
treatment of QED effects.

In our previous investigation \cite{yerokhin:12:lilike}, we performed a relativistic
configuration-interaction calculation of the $K\alpha$ transitions in lithium-like ions, including
iron. In that work, the QED effects were taken into account within the one-electron
screening-potential approximation. Theoretical results obtained in there agreed well with the
recent experimental data \cite{rudolph:13}, the theoretical precision being slightly better than
the experimental one.

In the present study, we apply this approach to a more complicated system, the beryllium-like iron.
For this ion, we calculate the energy levels of 5 lowest-lying and 18 core-excited states. When
combined with the data available for the helium- and lithium-like ions, accurate theoretical
results cover almost all experimentally observed Fe $K\alpha$ transitions in the region between
1.850 and 1.880 {\AA} \cite{beiersdorfer:93}.

The paper is organized as follows. In the next section, we give a brief outline of our computation
method with emphasis on new features as compared to the previous investigations. Sec.~\ref{sec:res}
then presents the results of our calculations and compares them with the previous theoretical and
experimental data. Relativistic units $\hbar=c=1$ and charge units $e^2/4\pi  = \alpha$ are used
throughout this paper.

\section{Method of calculation}

We perform our calculations of the energy levels in two steps. On the first step, we solve the
Dirac-Coulomb-Breit eigenvalue problem by means of the configuration-interaction (CI) method. On
the second step, we calculate the QED correction, which is then added to to the CI energy. The
nuclear recoil effect is small when compared to the total theoretical uncertainty and thus is taken
into account by means of the reduced mass prefactor (i.e., nonrelativistically and neglecting the
mass polarization).

\subsection{Dirac-Coulomb-Breit energy}

In relativistic quantum mechanics, the energy of an atom $E$ is given by the eigenvalue of the
Dirac-Coulomb-Breit (DCB) Hamiltonian $H_{\rm DCB}$,
\begin{eqnarray} \label{eql1}
    H_{\rm DCB}\,\Psi = E\, \Psi\,,
\end{eqnarray}
where $\Psi \equiv \Psi(PJM)$ is the $N$-electron wave function with given parity $P$, total
angular momentum $J$, and angular momentum projection $M$. The DCB Hamiltonian is conventionally
defined by
\begin{eqnarray}
    H_{\rm DCB} = \sum_i h_{\rm D}(i) + \sum_{i<j} \left[ V_{C}(i,j)+
    V_{B}(i,j)\right]\,,
\end{eqnarray}
where the indices $i,j = 1,\ldots,N$ numerate the electrons, $h_D$ is the one-particle
Dirac-Coulomb Hamiltonian, and $V_{C}$ and $V_B$ are the Coulomb and the frequency-independent
Breit parts of the electron-electron interaction, see, e.g., \cite{johnson:book} for details. It is
assumed that $H_{\rm DCB}$ acts in the space of the wave functions constructed from the
positive-energy eigenfunctions of some one-particle Dirac Hamiltonian (the so-called no-pair
approximation).

In the CI method, the eigenfunctions $\Psi(PJM)$ of Eq.~(\ref{eql1}) are represented by a (finite)
sum of the configuration-state functions (CSFs) with the same $P$, $J$, and $M$,
\begin{equation}\label{eq4}
  \Psi(PJM) = \sum_r c_r \Phi(\gamma_r PJM)\,,
\end{equation}
where $\gamma_r$ denotes the set of additional quantum numbers that determine the CSF. The CSFs are
constructed as linear combinations of antisymmetrized products of one-electron orbitals $\psi_n$.
In the present work, the one-electron orbitals $\psi_n$ are solutions of the frozen-core Dirac-Fock
equation.

The energy of the system is given by one of the roots of the secular equation
\begin{eqnarray}
    {\rm det} \bigl\{\lbr \gamma_r PJM|H_{\rm DCB}|\gamma_s PJM\rbr -E_r\,\delta_{rs}\bigr\} =
    0\,,
\end{eqnarray}
where ``det'' denotes the determinant of the matrix. The elements of the Hamiltonian matrix are
represented as linear combinations of one- and two-particle radial integrals (see, e.g.,
\cite{johnson:book}),
\begin{align}\label{eq7}
\lbr \gamma_r PJM| &\,  H_{\rm DCB}|\gamma_s PJM\rbr = \sum_{ab}
 d_{rs}(ab)\,I(ab)
  \nonumber \\
 + &\, \alpha \sum_k \sum_{abcd} v_{rs}^{(k)}(abcd)\,
  R_{k}(abcd)\,.
\end{align}
Here, $a$, $b$, $c$, and $d$ numerate the one-electron orbitals, $d_{rs}$ and $v^{(k)}_{rs}$ are
the angular coefficients \cite{grant:73:cpc,grant:76:jpb,gaigalas:01:cpc,gaigalas:02:cpc}, $I(ab)$
are the one-electron radial integrals, and $R_{k}(abcd)$ are the relativistic generalization of
Slater radial integrals \cite{johnson:88:b}.

Our implementation of the CI method has been described in the previous papers
\cite{yerokhin:08:pra:ra,yerokhin:08:pra,yerokhin:12:lilike}. We shall therefore discuss here only
those issues that are new to the present calculation. The first difficulty arises in the
calculation of the core-excited states, whose energies are well above the autoionization threshold,
i.e., above the continuum of the valence-excited states. For a large basis set we are using here,
the atomic states of interest turn out to be very far away from the lowest eigenvalue of the
Hamiltonian matrix. The numerical approach we were using previously for determining the eigenvalues
of a large matrix (the implementation of the Davidson algorithm by Stathopoulos and Froese Fischer
\cite{stathopoulos:94:cpc}) was suitable for the computation of just the lowest (highest) matrix
eigenvalues. In the present work, we use the Jacobi-Davidson algorithm as implemented within the
JDQZ package \cite{JDQZ,fokkema:98}. The JDQZ package, although significantly slower than the one
by Stathopoulos and Froese Fischer, was able to provide us with eigenvalues and the corresponding
eigenvectors around an arbitrary energy target far from the lowest eigenvalue.

Another new feature of the present calculations is the identification of computed levels in terms
of the nonrelativistic $LS$ coupling scheme. In our relativistic CI calculations we use the $jj$
coupling scheme, which is natural in the relativistic case. In order to compare the computed levels
with experiments and previous nonrelativistic calculations, we had to identify our calculated
levels within the $LS$ coupling scheme. In the case of the core-exited states of beryllium-like
ions such identification is not straightforward (as it was in the case of lithium-like ions),
because of the large number and high density of levels. In order to identify the levels, we
calculated the expectation values of the squares of the orbital momentum operator $L^2$ and the
spin operator $S^2$ with the eigenfunctions of the DCB Hamiltonian. The matrix elements of the
$L^2$ and $S^2$ operators are obtained as
\begin{align}
\lbr \gamma_r PJM| &\,  L^2 |\gamma_s PJM\rbr = \sum_{ab}
 d_{rs}(ab)\,I^{(L)}(ab)
  \nonumber \\
 + &\, \sum_{abcd} v_{rs}^{(1)}(abcd)\,
  R^{(L)}(abcd)\,,
\end{align}
and
\begin{align}
\lbr \gamma_r PJM| &\,  S^2 |\gamma_s PJM\rbr = \frac34\, \sum_{a}
 d_{rs}(aa)
  \nonumber \\
 + &\, \sum_{abcd} v_{rs}^{(1)}(abcd)\,
  R^{(S)}(abcd)\,,
\end{align}
where the angular coefficients $d_{rs}(ab)$ and $v_{rs}^{(k)}(abcd)$ are the same as in
Eq.~(\ref{eq7}) and the radial integrals $I^{(L)}(ab)$ and $R^{(L,S)}(abcd)$ are presented in
Appendix.

An essential part of the present calculation is the systematic estimation of the uncertainties of
the obtained theoretical predictions. For each atomic state of interest, we perform our CI
calculations with many (typically, about 20) different sets of configuration-state functions. From
these computations, we then deduce an estimate of how well our CI results were converged, by
analyzing the successive increments of the results obtained with the basis set being increased in
various directions.

\subsection{QED effects}

The QED effects are calculated in the present work by means of two different approaches. By
comparing the results from these approaches, we estimate the uncertainty of our treatment. The
first method is based on summing up the self-energy and vacuum-polarization QED corrections
calculated for each one-electron orbital in an effective screening potential. The total QED
correction for a given many-electron state is obtained by adding the QED contributions from all
one-electron orbitals, weighted by their fractional occupation numbers as obtained from the
eigenvectors of the CI calculation,
\begin{align}
\delta E_{\rm QED} = \sum_a q_a\,\bigl[\lbr a| \Sigma_{\rm SE}(\vare_a)|a\rbr
+  \lbr a| V_{\rm VP} |a\rbr  \bigr]\,.
\end{align}
Here, the index $a$ runs over all one-electron orbitals contributing to the given many-electron
state, $q_a$ is the occupation number of the one-electron orbital, $\Sigma_{\rm SE}$ is the
self-energy operator, $\vare_a$ is the Dirac energy of the one-electron state $a$, and $V_{\rm VP}$
is the vacuum polarization potential. This approximate treatment of QED corrections was used in our
previous work on lithium-like ions \cite{yerokhin:12:lilike} and similarly also by other authors,
in particular, for beryllium-like ions by Chen and Cheng \cite{chen:97}.

Our second method of evaluating the QED effects is based on the model QED operator $h^{\rm QED}$
formulated recently by Shabaev et al.~\cite{shabaev:13:qedmod} and implemented in the QEDMOD
Fortran package \cite{shabaev:14:qedmod}. The QEDMOD package is a tool that efficiently calculates
matrix elements of $h^{\rm QED}$ with the (bound-state) one-electron wave functions. In the present
work, we add the model QED operator to the DCB Hamiltonian, essentially modifying the one-electron
integrals $I(a,b)$ of Eq.~(\ref{eq7})  in our CI code by
\begin{align}
I(ab)\to I(ab) + \delta_{\kappa_a,\kappa_b}\,\lbr a|h^{\rm QED}|b\rbr\,,
\end{align}
where $\kappa_a$ denotes the relativistic angular quantum number of the state $a$. If either $a$ or
$b$ is a continuum state (i.e., $\max(\vare_a,\vare_b) > m$), the matrix element of $h^{\rm QED}$
is assumed to be zero. The QED correction to the energy level is then identified by taking the
difference of the CI eigenvalues with and without the $h^{\rm QED}$ operator. A comparison of the
results of these two approaches for evaluation of QED correction is presented in the next section.

\section{Results and discussion}
\label{sec:res}

In Table~\ref{tab:dcb} we present an example of our CI calculation of the Dirac-Coulomb-Breit
energy for the $1s2s2p^2\ $ $^3P_0$ state of beryllium-like iron. The various contributions in this
table are obtained by analyzing the results of calculations with 17 different sets of basis
functions. These basis sets are obtained by varying the number of partial waves included (i.e., the
largest value of the orbital momentum $l$ of one-electron orbitals), the size of the basis for each
partial wave, and by including or omitting the Breit interaction. By extending the basis set and
taking the differences of the results, we identify the contributions of individual partial waves
and check the stability of the results for each partial wave with regard to the number of basis
functions. The analysis is supplemented by estimating the tail of the expansion by polynomial
least-squares fitting of the increments in $1/l$.

The contribution of the triple excitations was found to be very small in all cases relevant for the
present work. We thus perform the main part of our calculations with single and double excitations
only, and estimate the contribution of the triple excitations separately within a smaller basis.
The partial-wave expansion was truncated at $l = 4$, with the contribution of the higher-$l$
multipoles being estimated by extrapolation. The typical size of the basis set was of about
$N=30,000$ functions. The results presented in Table~\ref{tab:dcb} are well converged with respect
to the number of partial waves as well as to the number of the basis functions. For higher excited
states, however, the convergence of the partial-wave expansion becomes slower and, more
importantly, the stability of the results with regard to the number of basis functions drops down.
The latter problem is associated with the interaction of the reference core-excited state with the
continuum of valence-exited states, which is difficult to describe accurately. For each atomic
state of interest, we perform a separate analysis of the convergence with different sets of basis
functions and estimate the uncertainty of the theoretical result based on this analysis.

In Table~\ref{tab:qed} we present results for the QED correction for selected states of
beryllium-like iron. As described above, the calculation is performed by two different methods,
namely, the model QED potential approach (QEDMOD) and the direct calculation of QED corrections in
a screening potential. In the latter case, we use two different screening potentials, the
core-Hartree (CH) potential and the localized Dirac-Fock (LDF) potential. The definition of these
potentials is the same as in our previous works \cite{yerokhin:12:lilike,yerokhin:08:pra}. Indeed,
we observe fair agreement between the QED corrections obtained by the different methods. In the
case of core-excited states, the difference between the results remains well within the 1\% range.
For the ground and valence-excited states, the deviation is noticeably larger, on the level of 2\%.
This is explained by a relatively large effect of screening of one $1s$ electron by another $1s$
electron, which is not well described by approximate methods.

As a final result for the QED correction we take the value  obtained by the model QED operator
approach. The uncertainty of this value was estimated by taking the maximal difference between the
three results. For the ground and the $^{3,1}P_1$ valence-excited states, our calculation can be
compared with the previous investigation by Chen and Cheng \cite{chen:97}, in which the
screening-potential approach with the Kohn-Sham screening potential was used. We observe that their
results are smaller and slightly outside of our error bars.

From our analysis, we conclude that the uncertainty of the theoretical predictions for the ground
and the valence-excited states of beryllium-like iron mainly arises from the uncertainty of the QED
treatment. A more rigorous QED calculation, similar to that for lithium-like ions
\cite{yerokhin:01:2ph,kozhedub:10}, would improve the theoretical accuracy for these states. In
contrast, for most of the core-excited states, the theoretical uncertainty comes both from the QED
effect and from the Dirac-Coulomb-Breit energy.

Table~\ref{tab:main} presents the calculated energy levels of beryllium-like iron. For the ground
$1s^22s^2\,^1S$ state, the total energy is listed, whereas  the {\em relative} energies (with
respect to the ground state) are given for the excited states. Our results are compared with the
NIST compilation based on experimental and theoretical data \cite{nist:13,shirai:00}, with the
relativistic CI calculation by Chen and Cheng \cite{chen:97}, as well as with the experimental
results \cite{denne:89,rudolph:13}. The estimated fractional accuracy of our theoretical energies
of the core-excited states is of about $2\times 10^{-5}$. The overall agreement with the NIST
compilation data is quite good, although we observe significant deviations up to ten times our
estimated uncertainty for a number of core-excited states. Notably, this is the case for the
$1s2s^22p\,^3P^o_1$ state, where excellent agreement is found with the recent measurement
\cite{rudolph:13}, well within the given error bars.

Finally, our theoretical results for the wavelengths of the $K\alpha$ transition lines of
beryllium-like iron are presented in Table~\ref{tab:trans}, together with the experimental results
\cite{beiersdorfer:93,seely:86:aj,rudolph:13}, the previous calculation by the $1/Z$ perturbation
theory \cite{shlyaptseva:98}, as well as the NIST spectral line compilation. The labelling of the
transition lines was taken from Ref.~\cite{beiersdorfer:93}. As seen from this table, different
transition lines are often very close to each other, so that small shifts in the theoretical
predictions might cause changes in the line ordering. In particular, our calculations lead to
re-assignment (interchange) of the E3 and E4 lines as well as the E8 and E9 lines, respectively.
The overall agreement of our calculations with the experimental results is very good, our results
being more accurate than the older astronomical measurements \cite{beiersdorfer:93,seely:86:aj} but
several times less precise when compared with the latest laboratory data \cite{rudolph:13}.

%%%%%%%%%%%%%%%%%%%%%%%%%%%%%%%%%%%%%%%%%%%%%%%%%%%%%%%%%%%%%%%%%%%%
\begin{table}
\caption{Contributions to the Dirac-Coulomb-Breit energy for the $1s2s2p^2\ $
$^3P_0$ state of beryllium-like iron, Fe$^{22+}$,
for inifitine nuclear mass, in atomic units. $SD$ denotes the contribution of
single and double excitations, $T$ denotes the contribution of the triple excitations.
$L$ is the maximal value of the orbital angular momentum quantum number of the
configuration-state functions.
\label{tab:dcb} }
\begin{ruledtabular}
  \begin{tabular}{l.}
  $L$ & \delta E   \\
    \hline\\[-5pt]
    \multicolumn{2}{l}{Coulomb, $SD$}\\
   1    &        -567.7x4335 \\
   2    &          -0.0x2467 \\
   3    &          -0.0x0351 \\
   4    &          -0.0x0082 \\
5$\ldots\infty$   &      -0.0x0058^a \\[5pt]
    \multicolumn{2}{l}{Breit, $SD$}\\
   1    &          0.0x4460 \\
   2    &         -0.0x0099 \\
   3    &         -0.0x0017 \\
   4    &         -0.0x0005 \\
5$\ldots\infty$  &     -0.0x0007^a \\    [5pt]
    \multicolumn{2}{l}{Coulomb, $T$}\\
   1    &         -0.0x0005 \\
    \hline\\[-5pt]
Total   & -567.7x2966\, (90)
  \end{tabular}
\end{ruledtabular}
$^a$ extrapolation.
\end{table}

\begin{table*}
\caption{QED corrections for beryllium-like iron, Fe$^{22+}$, in atomic units.
"QEDMOD" denotes the results obtained with the model QED potential, "LDF" labels the results obtained
with the localized Dirac-Fock potential, "CH" denotes results obtained with the core-Hartree potential.
"Final QED" denotes the final result for the QED correction with uncertainty.
\label{tab:qed} }
\begin{ruledtabular}
  \begin{tabular}{l......}
  Method & \multicolumn{1}{c}{$1s^22s^2\ $ $^1S_0$}
         & \multicolumn{1}{c}{$1s^22s2p\ $ $^3P_1^o$}
         & \multicolumn{1}{c}{$1s^22s2p\ $ $^1P_1^o$}
          &  \multicolumn{1}{c}{$1s2s^22p\ $ $^3P^o_1$}
         &  \multicolumn{1}{c}{$1s2s2p^2\ $ $^3P_0$}
         &  \multicolumn{1}{c}{$1s2s2p^2\ $ $^5P_3$}
          \\
    \hline\\[-5pt]
QEDMOD &  0.31x2\,25 & 0.29x6\,24 & 0.29x6\,60 &  0.17x6\,80 & 0.16x0\,77 & 0.16x1\,02  \\
LDF &     0.31x5\,30 & 0.29x4\,26 & 0.29x5\,02 &  0.17x6\,90 & 0.16x0\,28 & 0.16x1\,67   \\
CH  &     0.30x9\,17 & 0.29x3\,90 & 0.29x4\,68 &  0.17x6\,44 & 0.16x0\,03 & 0.16x1\,45   \\[5pt]
Final QED&    0.31x2\,2\,(30) & 0.29x6\,2\,(23) & 0.29x6\,6\,(19) &  0.17x6\,8\,(4) & 0.16x0\,8\,(8) & 0.16x1\,0\,(7)   \\
Ref.~\cite{chen:97}&
          0.30x8\,9       & 0.29x2\,9       & 0.29x3\,7 \\
  \end{tabular}
\end{ruledtabular}
\end{table*}

%%%%%%%%%%%%%%%%%%%%%%%%%%%%%%%%%%%%%%%%%%%%%%%%%%%%%%%%%%%%%%%%%%%%
\begingroup

\begin{table*}
\caption{Energy levels of beryllium-like iron Fe$^{22+}$, in Rydbergs,
1 Ry = 109\,737.315\,685\,39\,(55)~cm$^{-1}$. Separately listed are the
Dirac-Coulomb energy, the Breit correction, and the QED
correction.
The total energies are presented for the ground state, whereas
for all other states, the energies relative to the ground state are given.
The theoretical contributions are presented multiplied by the reduced mass
prefactor $\mu/m$, $1-\mu/m = 0.00000981$.
\label{tab:main}
}
%\squeezetable
\begin{ruledtabular}
\begin{tabular}{llcddddddd}
\multicolumn{2}{c}{Term} &
$J$ &
\multicolumn{1}{c}{$\ \ $Coulomb} &
\multicolumn{1}{c}{$\ \ \ \ \ $Breit} &
\multicolumn{1}{c}{$\ \ \ \ \ $QED} &
\multicolumn{1}{c}{$\ \ \ \ \ $Total} &
\multicolumn{1}{c}{$\ \ \ \ \ $NIST$^a$} &
\multicolumn{1}{c}{$\ \ \ \ \ $Other theory$^b$} &
\multicolumn{1}{c}{$\ \ \ \ \ $Experiment}
\\
\hline
\\[-5pt]
%
%            TERM                  J              COUL          BREIT       QED                TOTAL      NIST EXP
%
$1s^22s^2$           & $^1S$     & 0   &     -1625.6818 &    0.5181 &    0.6245 &      -1624.539 \, (6) &     &  -1624.547 \\[2pt]
$1s^22s2p$           & $^3P^o$     & 0   &         3.1650 &    0.0424 &   -0.0325 &          3.175 \, (9) &      3.173 \\
                     &           & 1   &         3.4632 &    0.0247 &   -0.0320 &          3.456 \, (9) &      3.455 &      3.455 & 3.4550\,(2)^c \\
                     &           & 2   &         4.3382 &   -0.0077 &   -0.0298 &          4.301 \, (9) &      4.299 \\[2pt]
$1s^22s2p$           & $^1P^o$     & 1   &         6.8856 &    0.0039 &   -0.0313 &          6.858 \, (9) &      6.856 &      6.858 & 6.8561\,(5)^c\\[2pt]
$1s2s^22p$           & $^3P^o$   & 1   &       485.6460 &   -0.4349 &   -0.2709 &        484.940 \, (19) &    484.786 && 484.934\,(5)^d  \\[2pt]
$1s2s^22p$           & $^1P^o$   & 1   &       487.9367 &   -0.4578 &   -0.2688 &        487.210 \, (23) &    487.200 && 487.208\,(5)^d\\[2pt]
$1s2s2p^2$           & $^5P$     & 1   &       486.7934 &   -0.3794 &   -0.3051 &        486.109 \, (7) &  \\
                     &           & 2   &       487.3533 &   -0.3851 &   -0.3035 &        486.665 \, (7) &  \\
                     &           & 3   &       487.9174 &   -0.4802 &   -0.3025 &        487.135 \, (7) &    487.200 \\[2pt]
$1s2s2p^2$           & $^3P$     & 0   &       490.1470 &   -0.4315 &   -0.3030 &        489.413 \, (7) &    489.414 \\
                     &           & 1   &       490.4131 &   -0.4291 &   -0.3029 &        489.681 \, (6) &  \\
                     &           & 2   &       491.4503 &   -0.4509 &   -0.3004 &        490.699 \, (7) &  \\[2pt]
$1s2s2p^2$           & $^3D$     & 1   &       490.8672 &   -0.3908 &   -0.3026 &        490.174 \, (7) &    490.260 \\
                     &           & 2   &       490.5592 &   -0.4403 &   -0.3034 &        489.816 \, (8) &    489.815 \\
                     &           & 3   &       490.9252 &   -0.5124 &   -0.3025 &        490.110 \, (8) &    490.134 \\[2pt]
$1s2s2p^2$           & $^3S$     & 1   &       492.0479 &   -0.3831 &   -0.3038 &        491.361 \, (9) &    491.401 \\[2pt]
$1s2s2p^2$           & $^1D$     & 2   &       492.7697 &   -0.3958 &   -0.3019 &        492.072 \, (7) &    492.053 \\[2pt]
$1s2s2p^2$           & $^3P$     & 0   &       492.1343 &   -0.3658 &   -0.3052 &        491.463 \, (8) &  \\
                     &           & 1   &       493.1155 &   -0.4072 &   -0.3022 &        492.406 \, (9) &  \\
                     &           & 2   &       493.6276 &   -0.4056 &   -0.3006 &        492.921 \, (8) &  \\[2pt]
$1s2s2p^2$           & $^1P$     & 1   &       494.4470 &   -0.4880 &   -0.3025 &        493.656 \, (10) &    493.743 \\[2pt]
$1s2s2p^2$           & $^1S$     & 0   &       494.8880 &   -0.3519 &   -0.2994 &        494.237 \, (11) &    494.381 \\
\end{tabular}
\end{ruledtabular}

$^a$ NIST atomic spectra data base \cite{nist:13} and Shirai et al. \cite{shirai:00},\\
$^b$ Chen and Cheng \cite{chen:97},\\
$^c$ Denne {\em et al.} \cite{denne:89},\\
$^d$ J. K. Rudolph {\em et al.} \cite{rudolph:13}. \\
\end{table*}
\endgroup

\begin{table*}
\caption{Transition line wavelengths of beryllium-like iron Fe$^{22+}$, in
  \AA. Line labeling is from Ref.~\cite{beiersdorfer:93}.
\label{tab:trans}
}
%\squeezetable
\begin{ruledtabular}
\begin{tabular}{cc....}
\multicolumn{1}{c}{Line} &
\multicolumn{1}{c}{Transition} &
\multicolumn{1}{c}{Present work} &
\multicolumn{1}{c}{Experiment} &
\multicolumn{1}{c}{Other theory$^d$} &
\multicolumn{1}{c}{NIST$^e$}
\\
\hline
\\[-5pt]
      & $1s^22s2p\,\,^3P_{1  } - 1s2s2p^2\,\,^3S_{1  }$  &   1.867x71 \,(3)  &                &   1.86x835      & \\[2pt]
 E2   & $1s^22s2p\,\,^3P_{2  } - 1s2s2p^2\,\,^1D_{2  }$  &   1.868x23 \,(2)  &                &   1.86x795      & \\[2pt]
 E4   & $1s^22s2p\,\,^1P_{1  } - 1s2s2p^2\,\,^1S_{0  }$  &   1.869x73 \,(3)  &                &   1.86x975      & 1.86x92 \\[2pt]
 E3   & $1s^22s^2\,\,^1S_{0  } - 1s2s^22p\,\,^1P^o_{1  }$  & 1.870x38 \,(9)  & 1.870x35 \,(8)^a &                &   1.87x05 \\
      &                                                &                   & 1.870x35 \,(11)^b\\
      &                                                &                   & 1.870x39\,(2)^c\\[2pt]
 E5   & $1s^22s2p\,\,^3P_{2  } - 1s2s2p^2\,\,^3S_{1  }$  &   1.870x95 \,(3)  &                &   1.87x155      &   1.87x08 \\[2pt]
      & $1s^22s2p\,\,^1P_{1  } - 1s2s2p^2\,\,^1P_{1  }$  &   1.871x96 \,(3)  &                &                &   1.87x14 \\[2pt]
 E6   & $1s^22s2p\,\,^3P_{1  } - 1s2s2p^2\,\,^3D_{1  }$  &   1.872x27 \,(2)  & 1.872x46 \,(35)^a&   1.87x255      &   1.87x24 \\
      &                                                &                   & 1.872x26 \,(23)^b\\[2pt]
 E7   & $1s^22s2p\,\,^3P_{0  } - 1s2s2p^2\,\,^3P_{1  }$  &   1.873x08 \,(2)  & 1.872x46 \,(35)^a&   1.87x285      & \\[2pt]
 E9   & $1s^22s2p\,\,^3P_{2  } - 1s2s2p^2\,\,^3P_{2  }$  &   1.873x50 \,(2)  & 1.873x47 \,(35)^a&   1.87x385      & \\[2pt]
 E8   & $1s^22s2p\,\,^3P_{1  } - 1s2s2p^2\,\,^3D_{2  }$  &   1.873x65 \,(2)  & 1.873x47 \,(35)^a&   1.87x355      &   1.87x36 \\
      &                                                &                   & 1.873x47 \,(15)^b\\[2pt]
 E10  & $1s^22s2p\,\,^1P_{1  } - 1s2s2p^2\,\,^3P_{2  }$  &   1.874x79 \,(2)  &                &   1.87x470      & \\[2pt]
 E11  & $1s^22s2p\,\,^3P_{2  } - 1s2s2p^2\,\,^3D_{1  }$  &   1.875x52 \,(2)  &                &   1.87x575      &   1.87x52 \\[2pt]
 E12  & $1s^22s2p\,\,^3P_{2  } - 1s2s2p^2\,\,^3D_{3  }$  &   1.875x77 \,(3)  & 1.875x74 \,(20)^a&   1.87x585      &   1.87x57 \\
      &                                                &                   & 1.875x52 \,(12)^b\\[2pt]
 E13  & $1s^22s2p\,\,^3P_{2  } - 1s2s2p^2\,\,^3D_{2  }$  &   1.876x91 \,(2)  &                &   1.87x675      & \\[2pt]
 E14  & $1s^22s2p\,\,^3P_{2  } - 1s2s2p^2\,\,^3P_{1  }$  &   1.877x43 \,(2)  &                &   1.87x715      & \\[2pt]
 E15  & $1s^22s2p\,\,^1P_{1  } - 1s2s2p^2\,\,^1D_{2  }$  &   1.878x07 \,(2)  & 1.878x12 \,(20)^a&   1.87x814      &   1.87x81 \\
      &                                                &                   & 1.877x98 \,(14)^b\\[2pt]
 E16  & $1s^22s^2\,\,^1S_{0  } - 1s2s^22p\,\,^3P^o_{1  }$  &   1.879x13 \,(7)& 1.879x33 \,(30)^a &        &   1.87x97 \\
      &                                                &                   & 1.879x57 \,(25)^b\\
      &                                                &                   & 1.879x16\,(2)^c\\    [2pt]
 E17  & $1s^22s2p\,\,^3P_{2  } - 1s2s2p^2\,\,^5P_{3  }$  &   1.887x33 \,(2)& 1.886x90 \,(35)^b & 1.887x45       &   \\
\end{tabular}
\end{ruledtabular}

$^a$ Beiersdorfer et al. \cite{beiersdorfer:93},\\
$^b$ J. F. Seely, et al. \cite{seely:86:aj}, with a
0.16 m{\AA} shift according to Ref.~\cite{beiersdorfer:93},\\
$^c$ J. K. Rudolph et al. \cite{rudolph:13}, \\
$^d$  A. S. Shlyaptseva et al. \cite{shlyaptseva:98},\\
$^e$ NIST atomic spectra data base \cite{nist:13} and Shirai et al. \cite{shirai:00}.
\end{table*}

\section{Conclusion}

In summary, we performed relativistic CI calculations of the energy levels of the ground, 4
valence-excited and 18 core-excited states in beryllium-like iron, Fe$^{22+}$.  Dirac-Coulomb-Breit
energies from extended CI calculations were combined with separately computed QED corrections. The
QED corrections were obtained by two approximate methods, the model QED operator approach and the
screening-potential approach. From the comparison of these two approaches we were able to estimate
the uncertainty of the overall QED shift. The uncertainty of the Dirac-Coulomb-Breit energies was
estimated on the basis of an analysis of the convergence of the CI results with respect to the
number of terms of the partial-wave expansion and the number of the one-electron basis functions.
The results obtained for the wavelengths of the $K\alpha$ transitions improve the previous
theoretical calculations and compare favourably with the experimental data.

%%%%%%%%%%%%%%%%%%%%%%%%%%%%%%%%%%%%%%%%%%%%%%%%%%%%%%%%%%%%%%%%%%%%%%%%%%%%%%%%%%%%%%
\section*{Acknowledgement}

We are grateful to Prof.~I.~I.~Tupitsyn for drawing our attention to the JDQZ package and for
useful advices on implementation of the CI method. The work reported in this paper was supported by
BMBF under Contract No.~05K13VHA.

%\bibliographystyle{../bibtex/phaip30}
%\bibliography{../bibtex/hfst}

\appendix

\section*{Appendix: Radial integrals for $\bm L^2$ and $\bm S^2$}

The radial integrals for the matrix elements of the $L^2$ operator are:
\begin{align}
I^{(L)}(ab) &\, = \delta_{\kappa_a,\kappa_b}\, \biggl[ l_a\,(l_a+1) \int_0^{\infty}dr\,r^2\,g_a(r)\,g_b(r)
  \nonumber \\ &\,
+ \overline{l}_a\,(\overline{l}_a+1) \int_0^{\infty}dr\,r^2\,f_a(r)\,f_b(r) \biggr]\,,
\end{align}
and
\begin{align}
R^{(L)}_{abcd}  &\, = (-1)^{l_a+l_b+j_c+j_d}\,2\,\Pi_{j_aj_bj_cj_d}\,{\cal R}_{ac}^{(L)}\,{\cal R}_{bd}^{(L)}\,,
\end{align}
with
\begin{align}
{\cal R}_{ac}^{(L)} &\, = \delta_{l_a,l_c}\biggl[ \Xi(l_a) \,
\SixJ{l_a}{1/2}{j_c}{j_a}{1}{l_a}\,
\int_0^{\infty}dr\,r^2\,g_a(r)\,g_c(r)
  \nonumber \\ &\,
  -\Xi(\overline{l}_a) \,
  \SixJ{\overline{l}_a}{1/2}{j_c}{j_a}{1}{\overline{l}_a}\,
\int_0^{\infty}dr\,r^2\,f_a(r)\,f_c(r)\biggr]\,.
\end{align}

The radial integrals for the $S^2$ operators are written in the similar form,
\begin{align}
R^{(S)}_{abcd}  &\, = (-1)^{l_a+l_b+j_a+j_b}\,2\,\Pi_{j_aj_bj_cj_d}\,{\cal R}_{ac}^{(S)}\,{\cal R}_{bd}^{(S)}\,,
\end{align}
with
\begin{align}
{\cal R}_{ac}^{(S)} &\, = \delta_{l_a,l_c}\biggl[ \Xi(1/2) \,
\SixJ{1/2}{l_a}{j_c}{j_a}{1}{1/2}\,
\int_0^{\infty}dr\,r^2\,g_a(r)\,g_c(r)
  \nonumber \\ &\,
  -\Xi(1/2) \,
  \SixJ{1/2}{\overline{l}_a}{j_c}{j_a}{1}{1/2}\,
\int_0^{\infty}dr\,r^2\,f_a(r)\,f_c(r)\biggr]\,.
\end{align}

The notations are as follows: $\Pi_{ab\ldots} = [(2a+1)(2b+a)\ldots]^{1/2}$, $\Xi(l) =
[l(l+1)(2l+1)]^{1/2}$, $\kappa_a$ is the relativistic momentum quantum number of the state $a$,
$j_a = |\kappa_a|-1/2$, $l_a = |\kappa_a+1/2|-1/2$, $\overline{l}_a = 2j_a-l_a$, and $g_a(r)$ and
$f_a(r)$ are the upper and the lower radial components of the one-electron Dirac wave function.

\end{document}